\documentclass[reprint,showpacs,showkeys,superscriptaddress,aps,pra]{revtex4-1}

\usepackage{graphicx}
\usepackage{epstopdf}
\usepackage{bm}
\usepackage{amssymb}
\usepackage{amsmath}
\usepackage{bbold}
\usepackage[colorlinks=true, linkcolor=blue]{hyperref}
\usepackage{ulem}
\usepackage{color, soul}
\usepackage{xcolor}
\usepackage{afterpage}
\usepackage{footmisc}
\usepackage{natbib}

\begin{document}

\title{Spin pumping and inverse spin Hall effect in CoFeB/IrMn heterostructures}
\author{Koustuv Roy}
\author{Abhisek Mishra}
\author{Pushpendra Gupta}
\author{Shaktiranjan Mohanty}
\author{Braj Bhusan Singh}

\address{Laboratory for Nanomagnetism and Magnetic Materials (LNMM), School of Physical Sciences, National Institute of Science Education and Research (NISER), HBNI, Jatni-752050, India}

 \author{Subhankar Bedanta}
 \email{sbedanta@niser.ac.in}
 \address{Laboratory for Nanomagnetism and Magnetic Materials (LNMM), School of Physical Sciences, National Institute of Science Education and Research (NISER), HBNI, Jatni-752050, India}
 \address{Center for Interdisciplinary Sciences (CIS), National Institute of Science Education and Research (NISER), HBNI, Jatni, 752050 India}

\begin{abstract}
 High spin to charge conversion efficiency is the requirement for the spintronics devices which is governed by spin pumping and inverse spin Hall effect (ISHE). In last one decade, ISHE and spin pumping are heavily investigated in ferromagnet/ heavy metal (HM) heterostructures. Recently antiferromagnetic (AFM) materials are found to be good replacement of HMs because AFMs exhibit terahertz spin dynamics, high spin-orbit coupling, and absence of stray field. In this context we have performed the ISHE in CoFeB/ IrMn heterostructures. Spin pumping study is carried out for $Co_{40}Fe_{40}B_{20} (12\ nm)/ Cu (3\ nm)/ Ir_{50}Mn_{50} (t\  nm)/ AlO_{x} (3\ nm)$ samples where \textit{t} value varies from 0 to 10 nm. Damping of all the samples are higher than the single layer CoFeB which indicates that spin pumping due to IrMn is the underneath mechanism. Further the spin pumping in the samples are confirmed by angle dependent ISHE measurements. We have also disentangled other spin rectifications effects and found that the spin pumping is dominant in all the samples. From the ISHE analysis the real part of spin mixing conductance (\textit{$g_{r}^{\uparrow \downarrow}$}) is found to be 0.704 $\pm$ 0.003 $\times$ $10^{18}$ $m^{-2}$.

\end{abstract}

%\pacs{}
\maketitle

\section{Introduction}
The generation and manipulation of the pure spin current is important for future technological devices and fundamental research as well \cite{1,2}. Generation of pure spin current via spin pumping \cite{5,6,7,8,9} is an efficient method in ferromagnetic (FM)/heavy metal (HM) systems. In this process the magnetization of FM is excited by a rf magnetic field through ferromagnetic resonance (FMR) and it generates the spin current ($\Vec{J}_s$) at the FM/HM interface, which is given by \cite{2,13}:	

\begin{equation}\label{equation1}
   {\Vec{J}_s}  = \frac{\hbar}{4 \pi}g_{r}^{\uparrow \downarrow}\hat{m} \times \frac{d \hat{m}}{dt}
\end{equation}

where $g_{r}^{\uparrow \downarrow}$ is the real part of spin mixing conductance and $\hat{m}$ is the unit vector of magnetization. The procession under time varying field can generate resultant dc current which diffuses towards FM/HM interfaces. It can be converted into transverse charge current and hence transverse voltage can be measured by inverse spin Hall effect (ISHE). The spin to charge current conversion efficiency mostly depends on SOC of HM and its conductivity \cite{2,14}. However there are only a few HMs in the periodic table and usually they are quite expensive. On the other hand recent works show antiferromagnets (AFM) exhibit high SOC and efficient transfer of spin angular momentum via AFM spin waves \cite{zhang2014spin, singh2020inverse,singh2020large}. In last few decades FM/AFM bilayers have been heavily used in spintronics device application due to the exchange bias property \cite{nogues1999exchange}. In addition to that the ISHE study on this FM/AFM bilayers can give rise to new scope in spintronics for future device applications.
 The collinear AFMs (e.g., $Mn_2Au$, IrMn, PtMn, FeMn, PdMn etc.) have drawn attention in this field for exhibiting high charge to spin current conversion efficiency \cite{singh2020large, zhang2014spin, zhou2019large}. It has been shown that among various CuAu-I type AFMs, IrMn is one of the potential candidate for ISHE studies. Most of the ISHE works on IrMn have been performed with crystalline FM materials like NiFe (Py) \cite{hao2019magnetization,wu2015observation,frangou2016enhanced}. There are very few reports of spin pumping with CoFeB/IrMn bilayers \cite{jhajhria2018influence,tu2017anomalous, ma2018correlation}. Jhajharia \textit{et. al.} have studied spin pumping and inter-layer exchange coupling via damping analysis in CoFeB/IrMn/CoFeB trilayer structure for variable IrMn thickness \cite{jhajhria2018influence}. Tu \textit{et. al.} have estimated the Nernst coeffecient via anomalous Nernst effect \cite{tu2017anomalous} in a IrMn/CoFeB system with perpendicular magnetic anisotropy. Further Ma \textit{et. al.} have corelated the spin mixing conductance and Dzyaloshinskii-Moriya interaction via Brillouin light-scattering
measurements \cite{ma2018correlation}. However to the best of our knowledge the detailed ISHE measurement in IrMn/CoFeB system has not been reported. In this work, we have systematically studied the ISHE on CoFeB/Cu/IrMn heterostructures. The crystallinity of the FM also plays a role in spin transport phenomena as it directly affects the interface of the FM/AFM. We have introduced a spacer Cu(3 nm) layer to improve the growth of IrMn. The lower thickness of Cu layer in comparison to its spin diffusion length ($\sim$350 nm) \cite{jedema2003spin} allows unperturbed spin current propagation from CoFeB to IrMn.

\section{Experimental Details}
Table \ref{table1} shows the sample structure of our study. All the samples were prepared on Si(100) substrates using magnetron sputtering in a vacuum system with base pressure $\sim$ 5 $\times$ $10^{-8}$ mbar. R1 and R2 are called the reference samples. The thickness of the films were evaluated using x-ray reflectivity (XRR) (data not shown) and transmission electron microscope(TEM) imaging. Crystalline quality of the thin films is also characterized by grazing incidence X-ray diffraction (GIXRD) and TEM imaging. The wave length of x-rays is 0.1541 nm. The DC magnetometry was performed by superconducting quantum interference device (SQUID) based magnetometer.

Ferromagnetic resonance (FMR) measurements have been performed in the frequency range 5-16 GHz on a coplanar wave guide in the flip-chip manner \cite{25,26}. ISHE measurements have been performed by connecting a nanovoltmeter over two ends of the sample (sample size: 3 mm $\times$ 2 mm). The details of the ISHE set-up can be found elsewhere \cite{singh2020inverse,singh2020large,26,singh2020high,SinghHighspin2021,gupta2021simultaneous,sahoo2020spin,27}.

\section{Results and Discussion}

\subsection*{Crystalline quality}
\begin{figure*}[htb]
	\centering
	\includegraphics[width=0.75\textwidth]{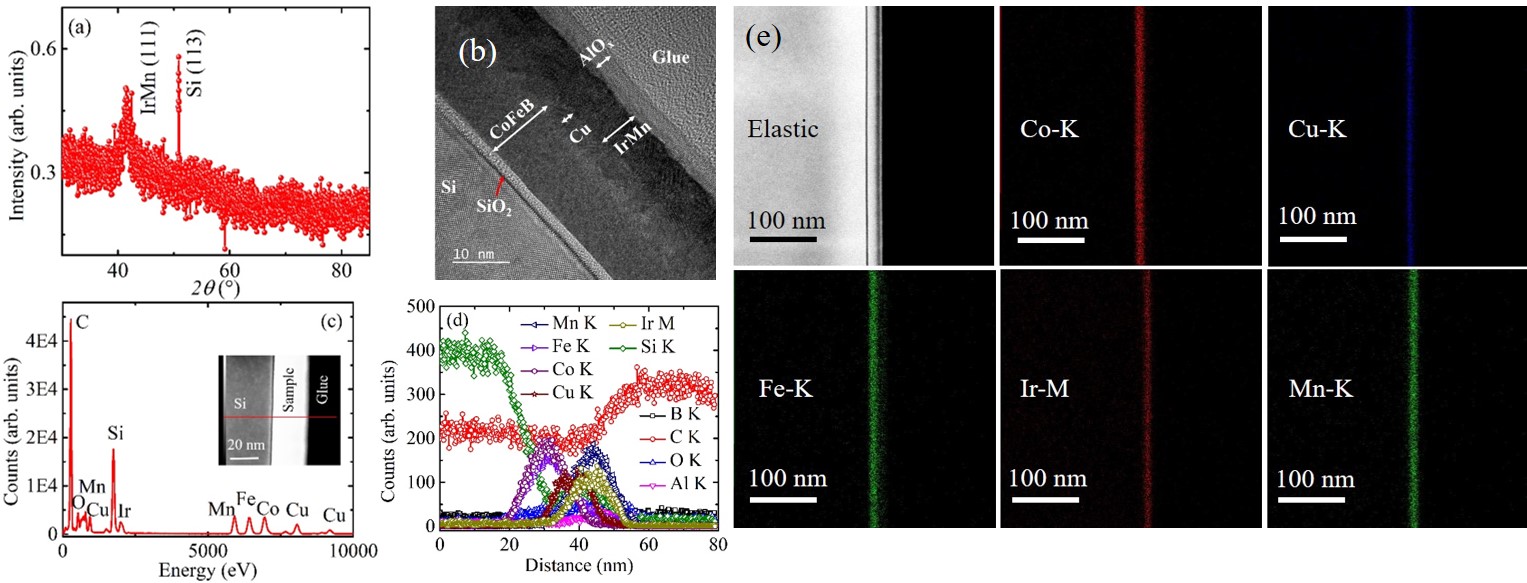}
	\caption{(a) GIXRD pattern, (b) high resolution cross-sectional TEM image, (c) line scan EDX (inset image shows the selected area),(d) corresponding EDX scan profile of sample S6. (e) Elemental mapping using Energy-filtered transmission electron microscopy (EFTEM) for S6 at different energy edges.}
	\label{TEM}
\end{figure*}

 Fig. \ref{TEM}(a) shows the GIXRD pattern of the sample S6. We could clearly see the XRD peaks at 41.4$^{\circ}$ correspond to IrMn(111). The CoFeB has been grown in the amorphous form, therefore it does not show any XRD peaks.

The fig.\ref{TEM}(b) shows the TEM image of sample S6. It clearly shows the amorphous nature of the CoFeB and IrMn has been grown in polycrystalline phase. Further, we have performed line scan energy dispersive x-ray analysis (EDX) of the sample which are shown in fig.\ref{TEM} (c,d). The EDX result confirms the existence of all the elements in the sample structure. Fig. \ref{TEM}(e) shows the elemental mapping of S6 where different elements are represented in different colours.

\begin{table}[]
\caption{Samples studied for the present work (The numbers in the bracket are in the units of nm).}
        \centering
\begin{tabular}{|c|c|}
\hline
R1 & CoFeB(12)                                                           \\ \hline
R2 & CoFeB(12)/Cu(3)                                                     \\ \hline
S1 & CoFeB(12)/Cu(3)/$AlO_{x}$(3)          \\ \hline
S2 & CoFeB(12)/Cu(3)/IrMn(0.5)/$AlO_{x}$(3)  \\ \hline
S3 & CoFeB(12)/Cu(3)/IrMn(0.75)/$AlO_{x}$(3) \\ \hline
S4 & CoFeB(12)/Cu(3)/IrMn(2)/$AlO_{x}$(3)    \\ \hline
S5 & CoFeB(12)/Cu(3)/IrMn(5)/$AlO_{x}$(3)   \\ \hline
S6 & CoFeB(12)/Cu(3)/IrMn(10)/$AlO_{x}$(3) \\ \hline
\end{tabular}
	\label{table1}
\end{table}

\subsection*{Magnetic Damping}

The Gilbert damping($\alpha$) of the samples was evaluated using frequency dependent FMR spectra (see fig. 1 and its corresponding explanation in the supplementary information) \cite{28,29}. The values of $\alpha$ for the samples S1-S6 are 0.0085, 0.0102, 0.0101, 0.0100, 0.0103, 0.0099 with the error bar $\pm$ 0.0001, respectively. It is observed that the $\alpha$ gets enhanced by significant margin in the bilayer samples S2-S6 due to presence of the IrMn layer. We found that the spacer Cu layer has negligible effect on $\alpha$ of CoFeB (for detail analysis refer to the supplementary information).

It is observed that the value of $\alpha$ for the sample S2 is maximum in which the IrMn thickness is 0.5 nm. For samples with higher IrMn thickness, the change in $\alpha$ value is negligible. This behaviour can be explained by considering the spin diffusion length of IrMn ($\sim$ 0.7 nm) \cite{zhang2014spin}. In case of dominant spin pumping behaviour over the other interface effects, it is usually observed that the $\alpha$ value saturates at the NM thickness equals to it's spin diffusion length \cite{conca2016study}. 

\subsection*{Inverse spin Hall effect measurement}

Damping analysis shown above indicates the presence of spin pumping mechanism in the samples. In order to confirm the spin pumping and quantitative analysis of the spin current generation in our system, we have performed ISHE measurements on all the samples. The measurements are carried out at 7 GHz frequency and +15 dBm power.

\begin{figure}[]
	\centering
	\includegraphics[width=0.5\textwidth]{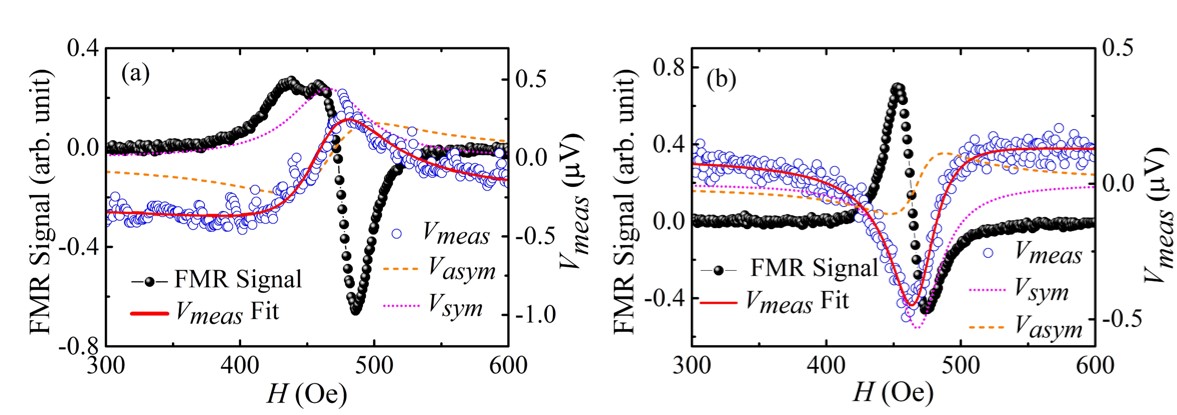}
	\caption{Voltage (\textit{$V_{meas}$}) measured across the sample with applied magnetic field along with FMR signal for sample S6 at the $\phi$ values of (a) 0$^{\circ}$, (b) 180$^{\circ}$. Experimental data is represented as open symbol. Solid lines are the fit to the experimental data using Eq. (\ref{q7}). Short dotted and dash lines are the symmetric ($V_{sym})$ and anti-symmetric ($V_{asym})$ components of the voltage.}
	\label{fig3}
\end{figure}

\begin{table*}[t]
\caption{Fitted parameters from $\phi$ dependent voltage measurements for the samples S2-S6}

\centering
\begin{tabular}{ccccc} \label{table2}

Sample & $V_{sp}$(V)$\times$$10^{-6}$ & $V_{AHE}$(V)$\times$$10^{-6}$ & $V_{AMR}^{\perp}$(V)$\times$$10^{-6}$ & $V_{AMR}^{||}$(V)$\times$$10^{-6}$ \\ \hline
S2     & 1.14 $\pm$ 0.03              & -0.43 $\pm$ 0.01              & 0.86 $\pm$ 0.01                       & 0.14 $\pm$ 0.02                    \\ \hline
S3     & 1.49 $\pm$ 0.04              & -0.74 $\pm$ 0.02              & 1.00 $\pm$ 0.01                       & 0.23 $\pm$ 0.05                    \\ \hline
S4     & 1.47 $\pm$ 0.04              & -0.75 $\pm$ 0.02              & 0.94 $\pm$ 0.02                       & 0.33 $\pm$ 0.02                    \\ \hline
S5     & 1.21 $\pm$ 0.02              & -0.18 $\pm$ 0.03              & 0.67 $\pm$ 0.01                       & 0.10 $\pm$ 0.02                    \\ \hline
S6     & 1.24 $\pm$ 0.04              & -0.65 $\pm$ 0.02              & 0.81 $\pm$ 0.01                       & 0.09 $\pm$ 0.02                    \\ 
\end{tabular}

\end{table*}

The measured dc voltage ($V_{meas}$) may have contributions from spin pumping, anisotropic magnetoresistance (AMR) and anomalous Hall effect (AHE). Angle dependent voltage measurement have been performed to disentangle these spurious effects. Fig. \ref{fig3} shows \textit{$V_{meas}$} (open blue circles) versus \textit{H} along with FMR signal (solid black circles) for sample S6 at the angles $\phi$= 0$^{\circ}$ (a) and 180$^{\circ}$ (b). The $\phi$ denotes the angle between the perpendicular direction of applied DC magnetic field (\textit{H}) and measured voltage direction. It should be noted that the pumped spin current is invariant upon the sample rotation in the measurement. Upon the change in polarity of the voltmeter electrodes, the measured voltage should change sign in case of spin pumping in the samples. We found that the measured voltage ($V_{meas}$) is changing sign with the 180$^{\circ}$ rotation of the sample which is a clear indication of the presence of spin pumping in the sample. It should also be noted that the $V_{meas}$ goes to zero value when the $\phi$ $\sim$ 90$^{\circ}$ (data not shown). According to the basic mechanism of electron spin scattering in ISHE, it happens because the spin accumulation is almost zero when the measured voltage direction is parallel to applied magnetic field. The maximum spin accumulation happens in the perpendicular direction to the applied magnetic field. In order to, quantify spin pumping contribution, the $\it{V_{meas}}$ versus $\it{H}$ data for the samples S6 (Fig. \ref{fig3}) were fitted with Lorentzian equation\cite{33}, which is given by:

\begin{equation}\label{q7}
\begin{aligned}
    V_{meas} = V_{sym} \frac{(\Delta H)^2}{(H-H_r)^2+(\Delta H)^2}+ \\
    V_{asym} \frac{2 \Delta H (H - H_r)}{(H-H_r)^2+(\Delta H)^2}
    \end{aligned}
\end{equation}
\begin{figure}[]
	\centering
	\includegraphics[width=0.33\textwidth]{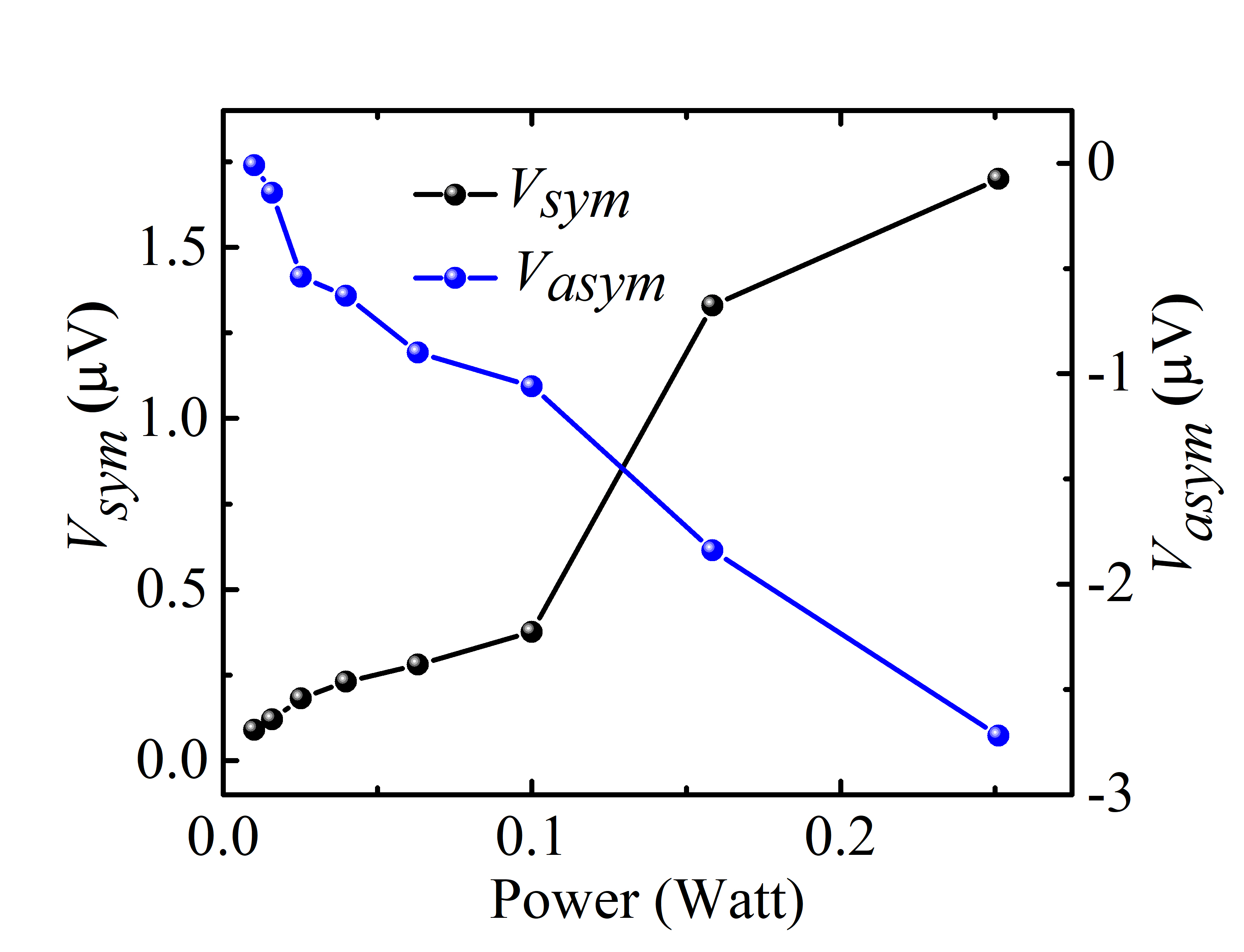}
	\caption{\textit{$V_{sym}$} and \textit{$V_{asym}$} contribution with the variation of \textit{rf} magnetic field power for the sample S6.}
	\label{fig4}
\end{figure}

where \textit{$V_{sym}$} and \textit{$V_{asym}$} are the symmetric and anti-symmetric Lorenzian components of the measured signal. Solid red lines are fits to the experimental data. The $V_{sym}$ consists of spin pumping, AMR, and AHE  effects, whereas the AMR and AHE are only contribute to \textit{$V_{asym}$}. Fig. \ref{fig3} also represents the components \textit{$V_{sym}$} (dotted line) and \textit{$V_{asym}$} (dashed line) plotted separately for the sample S6.

\begin{figure}[htb]
	\centering
	\includegraphics[width=0.5\textwidth]{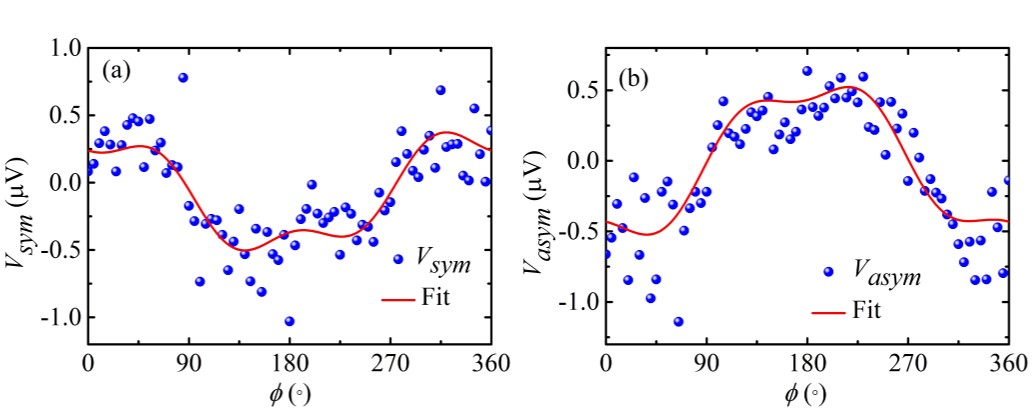}
	\caption{Angle ($\phi$) dependent (a) $V_{sym}$ and (b) $V_{asym}$ measurements for sample S6.}
	\label{fig5}
\end{figure}

In order to quantify the spin pumping and other rectification effects (viz. AMR and AHE), in-plane angle dependent ISHE measurements of \textit{$V_{meas}$} were performed at the interval of 5\textit{$^{\circ}$} angle for all the samples (fig.\ref{fig5} shows for S6). This method is an established one to decouple the individual components from the measured voltage  \cite{30,34,35}. Harder \textit{et.al.}  \cite{36} has considered the rectification effects i.e., the AHE contribution due to the FM layer, perpendicular AMR (\textit{$V_{asym/sym}^{AMR \perp}$}) and parallel AMR (\textit{$V_{asym/sym}^{AMR ||}$}) to the applied \textit{rf} field. The relation between these rectification effects and $V_{meas}$ are given in the below equations \ref{q8_new} and \ref{q9_new} \cite{34}. It should be noted that the FMR measurements were performed in the quasi-TEM mode, where the rf magnetic and electric fields are perpendicular to each other. Therefore, the phase angle ($\Phi$) between the rf electric and magnetic field is $90^{\circ}$.

\begin{equation}\label{q8_new}
\begin{aligned}
 V_{asym}= V_{AHE} cos(\phi + \phi_0) + \\
    V_{asym}^{AMR \perp} cos 2(\phi + \phi_0)cos(\phi + \phi_0)+ 
   \\ V_{asym}^{AMR ||}sin2(\phi + \phi_0)cos(\phi+\phi_0)
   \end{aligned}
 \end{equation}

\begin{equation}\label{q9_new}
\begin{aligned}
    V_{sym}= V_{sp}cos^3(\phi + \phi_0)+\\
    + V_{sym}^{AMR \perp} cos 2(\phi + \phi_0)cos(\phi+ \phi_0)\\
    + V_{sym}^{AMR ||}sin2(\phi + \phi_0)cos(\phi+\phi_0)
    \end{aligned}
\end{equation}

\textit{$V_{sp}$} and \textit{$V_{AHE}$} correspond to the spin pumping and AHE contributions, respectively. The extra factor $\phi_{0}$ takes care of the misalignment of sample positioning in defining the $\phi$ value during the measurement \cite{SinghHighspin2021, gupta2021simultaneous}.
The \textit{$V_{asym}^{AMR \perp,||}$} and \textit{$V_{sym}^{AMR \perp,||}$} are evaluated by fitting the angle dependent data of $V_{asym}$ and $V_{sym}$ to equations \ref{q8_new} and \ref{q9_new}. The best fits are shown in fig \ref{fig5}. Various components such as $V_{sp}$, $V_{AHE}$, $V_{AMR}$ etc. for all the samples are listed in Table \ref{table2}.
It should be noted that the sample S1 which is without IrMn does not show any ISHE signal (data not shown). It is expected as Cu does not have any high SOC. This fact confirms that the spin pumping mechanism is occurring in the samples S2-S6 solely because of IrMn layer.

Table \ref{table2} clearly shows that the \textit{$V_{sp}$} is dominating over AMR and AHE effects in all the samples.  Fig. \ref{fig4} shows the \textit{rf} magnetic field power dependent $V_{sym}$ and $V_{asym}$ components for the sample S6. With the increase in power, the spin angular momentum transfer from FM to NM increases linearly. The linear behaviour in fig. \ref{fig4} is another signature of the spin pumping mechanism in the CoFeB/IrMn samples.

It is observed that \textit{$V_{sp}$} is maximum for sample S3. It is known that the spin pumping will be maximum at the thickness of NM equal to its spin diffusion length. The spin diffusion length ($\lambda_{IrMn}$) is reported to be 0.7 nm \cite{zhang2014spin}. The thickness of IrMn in sample S3 is nearly same to the value of $\lambda_{IrMn}$, therefore, we have observed highest value of \textit{$V_{sp}$} compared to other samples. It has been shown that the \textit{$V_{sp}$} depends upon the conductivity of the HM \cite{ando2011inverse}. At higher thickness of IrMn, the conductivity of IrMn increases. This leads to decrease in \textit{$V_{sp}$} value for samples S5 and S6. In addition to this, the back spin pumping also dominates beyond the spin diffusion length of HM due to the spin flip mechanism at this thickness limit. Thus the decrease in \textit{$V_{sp}$} value for S5 and S6 is possibly a cumulative contribution of conductivity as well as the spin back flow.

\begin{figure}[ht]
	\centering
	\includegraphics[width=0.33\textwidth]{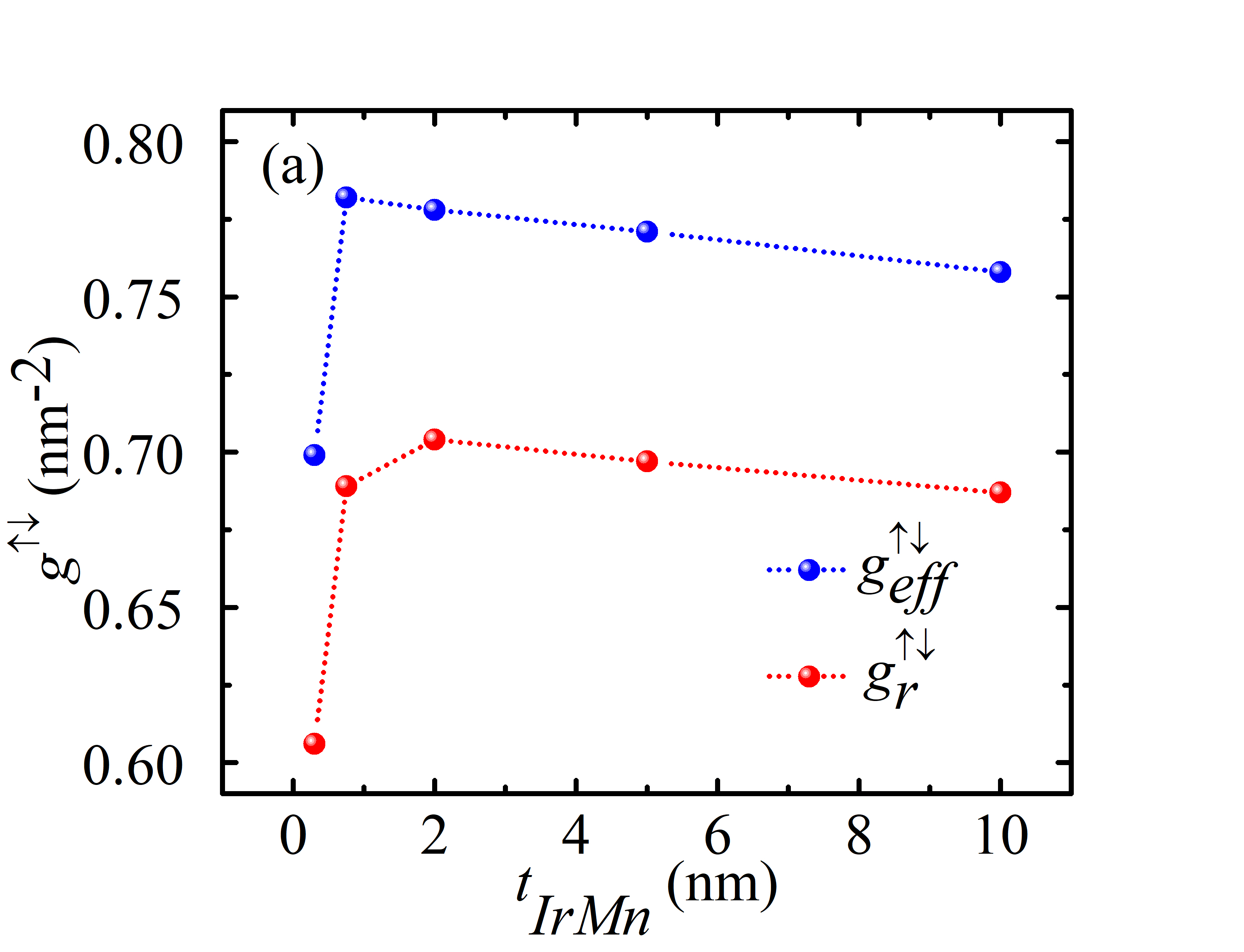}
	\caption{$g_{\it{eff}}^{\uparrow \downarrow}$, \textit{$g_{\it{r}}^{\uparrow \downarrow}$} with the variation of IrMn thickness.}
	\label{fig6}
\end{figure}

Effective spin mixing conductance (\textit{$g_{\it{eff}}^{\uparrow \downarrow}$}) is another parameter which defines the efficiency of spin current propagation from FM to NM. \textit{$g_{\it{eff}}^{\uparrow \downarrow}$}  was calculated by the following expression \cite{2}:

\begin{equation}\label{q11}
 g_{eff}^{\uparrow\downarrow}=\frac{\Delta\alpha 4\pi M_{s}t_{CoFeB}}{g\mu_{B}} 
\end{equation}

where \textit{$\Delta\alpha$}, \textit{$t_{CoFeB}$}, \textit{$\mu_{B}$}, \textit{g} are the change in the $\alpha$ due to the presence of IrMn, the thickness of CoFeB layer, Bohr magneton, Lande g- factor, respectively. Figure \ref{fig6} shows the plot between \textit{$g_{\it{eff}}^{\uparrow \downarrow}$} and IrMn thickness. The real part (\textit{$g_{\it{r}}^{\uparrow \downarrow}$}) of  \textit{$g_{\it{eff}}^{\uparrow \downarrow}$} mainly contributes to the spin transport mechanism. The \textit{$g_{\it{r}}^{\uparrow \downarrow}$} value can be evaluated by the following expression\cite{jeon2018spin,rogdakis2019spin}:

\begin{equation}
g_{r}^{\uparrow\downarrow}=g_{eff}^{\uparrow\downarrow}[1+\dfrac{g_{eff}^{\uparrow\downarrow}\rho_{IrMn}\lambda_{IrMn}e^2}{2\pi\hbar\tanh[\dfrac{t_{IrMn}}{\lambda_{IrMn}}]}]^{-1} 
\end{equation}

where $\rho_{IrMn}$, $t_{IrMn}$ are the resistivity and thickness of IrMn, respectively. In order to calculate \textit{$g_{\it{r}}^{\uparrow \downarrow}$}, the $\lambda_{IrMn}$ is taken to be 0.7 nm \cite{zhang2014spin}. The \textit{$g_{\it{r}}^{\uparrow \downarrow}$} values for samples S2-S6 are found to be 0.606, 0.690, 0.704, 0.696, 0.687 $nm^{-2}$ with a error bar of $\sim \pm$ 0.003 $nm^{-2}$. It is also to be noted that our \textit{$g_{r}^{\uparrow \downarrow}$} values are one order less than the previously reported values with IrMn \cite{zhang2014spin,frangou2016enhanced, holanda2020magnetic,ma2018correlation}. The relatively lower value of \textit{$g_{r}^{\uparrow \downarrow}$} confirms that the spin current flow efficiency is not so good in the studied heterostructures. It is probably due to the poor interfaces in CoFeB/Cu/IrMn heterostructures as seen in the TEM studies shown in fig. \ref{TEM}(b). Spin interface transparency (\textit{T}) is the parameter which defines the spin current propagation efficiency at the FM/NM interface.\textit{T} value is calculated from the following expressions \cite{43}:

\begin{equation}\label{q13}
    T = \frac{g_{r}^{\uparrow \downarrow} tanh(\frac{t_{IrMn}}{2 \lambda_{IrMn}})}{g_{r}^{\uparrow \downarrow} coth(\frac{t_{IrMn}}{\lambda_{IrMn}})+\frac{h\sigma_{IrMn}}{2e^2\lambda_{IrMn}}}
\end{equation}

where \textit{$\sigma_{IrMn}$} is the conductivity of IrMn layer. The maximum \textit{$g_{\it{r}}^{\uparrow \downarrow}$} is found to be in sample S4. The \textit{$\sigma_{IrMn}$} is found to be 0.18 $\times$ $10^{6}$  $\Omega^{-1}$.$m^{-1}$ using the four probe technique. For S4 (\textit{$t_{IrMn}$}= 2 nm), \textit{T} is evaluated to be 0.14 $\pm$ 0.01 by Eq.\ref{q13}, which is lesser than the values reported in the literature for Pt based systems  \cite{43}.

\section{Conclusion}
We presented the spin pumping and inverse spin Hall effect study in a series of CoFeB/ IrMn heterostructures. The IrMn has been grown in AFM phase. Angle dependent study was performed to disentangle all the spin rectification effects. Strong spin pumping contribution is found in the samples. The spin pumping voltage saturates at the spin diffusion length of the IrMn thickness. The maximum \textit{$g_{r}^{\uparrow \downarrow}$} is evaluated to be 0.704 $\times 10^{18}$ $m^{-2}$ for sample S4. Spin transparency also found to be 0.14 which indicates that Cu is not a good spacer layer for IrMn and CoFeB. But, the prominent spin pumping voltage predicts that IrMn is a good replacement of HM in spintronic device applications.

\section*{Acknowledgments}
The authors acknowledge DAE and DST, Govt. of India, for the financial support for the experimental facilities. KR and PG thank CSIR and UGC for their JRF fellowships, respectively. BBS acknowledges DST for INSPIRE faculty fellowship.

\section*{References}

\bibliographystyle{iopart-num}
\bibliography{reference}

\end{document}